\documentclass[aip,amsmath,amssymb,reprint, floatfix]{revtex4-1}
\usepackage[english]{babel} 
\usepackage{tikz}
\usepackage{orcidlink}
\usetikzlibrary{shapes, arrows, positioning}
\usepackage{graphicx}
\usepackage{dcolumn}
\usepackage{bm}
\usepackage[utf8]{inputenc}
\usepackage[T1]{fontenc}
\usepackage{mathptmx}
\usepackage{etoolbox}
\usepackage{float}
\usepackage{tikz}
\usepackage{multirow}
\usetikzlibrary{shapes.geometric, arrows, positioning}

\makeatletter
\def\@email#1#2{%
 \endgroup
 \patchcmd{\titleblock@produce}
  {\frontmatter@RRAPformat}
  {\frontmatter@RRAPformat{\produce@RRAP{*#1\href{mailto:#2}{#2}}}\frontmatter@RRAPformat}
  {}{}
}%
\makeatother
\begin{document}
\selectlanguage{english} 

\preprint{AIP/123-QED}

\title[RinQ: Towards predicting central sites in proteins on current quantum computers]{RinQ: Towards predicting central sites in proteins on current quantum computers}
\author{Shah Ishmam Mohtashim \orcidlink{0000-0002-7382-6466}}
 \affiliation{Department of Chemistry, North Carolina State University, Raleigh, NC, 27606, USA}
  \affiliation{Department of Chemistry, Purdue University, West Lafayette, IN 47907, USA}
  \email{smohtas@ncsu.edu}


\date{\today}

\begin{abstract}
We introduce RinQ, a hybrid quantum-classical framework for identifying functionally critical residues in proteins by formulating centrality detection as a Quadratic Unconstrained Binary Optimization (QUBO) problem. Protein structures are modeled as residue interaction networks (RINs), and the QUBO formulations are solved using D-Wave’s simulated annealing. Applied to a diverse set of proteins, RinQ consistently identifies central residues that closely align with classical benchmarks, demonstrating both the accuracy and robustness of the approach.
\end{abstract}

\maketitle

\section{Introduction}
Proteins, the fundamental building blocks of life, rely on finely tuned interactions between amino acid residues to maintain structural integrity and enable dynamic functionality. Identifying residues that play central roles in these interactions---often referred to as ``hotspots" or ``active sites"---is a longstanding goal in structural biology, with implications spanning protein engineering, drug discovery, and understanding the molecular basis of disease. Conventionally, classical network analysis techniques have been employed to pinpoint such critical sites within proteins, typically through measures such as degree, betweenness, closeness, and eigenvector centrality \cite{Negre2018,TAYLOR2013e201302006,doi:10.1021/acs.jcim.8b00146}.
In recent years, advances in quantum computing and quantum-inspired algorithms have opened new avenues for tackling combinatorial optimization problems inherent in biological systems \cite{fingerhuth2018quantumalternatingoperatoransatz, Perdomo-Ortiz2012-do}.

In this work, we introduce RinQ (Residue Interaction Network Quantum engine), a hybrid quantum-classical framework designed to identify central residues in proteins through a Quadratic Unconstrained Binary Optimization (QUBO) formulation of centrality measures. RinQ leverages residue interaction networks (RINs)---graph-theoretic abstractions of protein structures---as its foundational model. In this work, it is shown that this approach can consistently identify key residues, matching classical benchmarks, thus setting the stage for future exploration of quantum computational techniques in this area. Our study applies RinQ across a diverse set of proteins, from small peptide fragments to regulatory proteins of significant biological importance. 

Through systematic testing, we demonstrate that RinQ not only replicates classical eigenvector centrality results but also captures alternative network characteristics through Estrada centrality, providing a comprehensive picture of residue importance. In addition to providing the theoretical underpinnings of our approach, we present an extensive comparative analysis across multiple protein structures, detailing the implementation of quantum and classical solvers. We have also briefly analyzed the computational complexity of this formulation. By addressing the practical challenges of protein structure-function analysis, this work represents a vital contribution to current efforts to bridge quantum computing and bioinformatics \cite{qBinf24, shehab24, aplQ, trev_ieee, khatami22}. 


\section{Theory}

In RINs, nodes represent amino acid residues, and edges represent significant interactions between them, typically based on spatial proximity or energetic criteria \cite{Di_Paola2013-rr}. This transformation enables the application of computational network algorithms to extract biologically meaningful insights \cite{K2005-ki}. To construct a RIN, a protein structure is typically obtained from the Protein Data Bank (PDB), and an interaction graph is created by connecting residues whose C$\alpha$ atoms lie within a specified cutoff distance. We took 8 \AA  as our cutoff. The use of an 8~\AA{} cutoff in constructing Residue Interaction Networks (RINs) is grounded in both geometric and empirical considerations. Geometrically, this distance corresponds approximately to two peptide bond lengths, making it a meaningful upper bound for capturing inter-residue interactions that are biologically significant. As emphasized by Di Paola \textit{et al.}~(2013), ``the upper threshold of 8~\AA{}, commonly introduced in the analysis of RINs, roughly corresponds to two peptide bond lengths. Most authors consider only an upper threshold (around 8~\AA{}) to cut off negligible interactions''~\cite{DiPaola2013}. This convention is designed to retain functionally relevant non-covalent contacts---such as van der Waals interactions and hydrogen bonds---while excluding spurious long-range distances that do not contribute meaningfully to the structural integrity or communicability of the protein. Consequently, the 8~\AA{} threshold is widely adopted across studies of residue interaction networks (RINs) and rigidity-based residue geometry networks~\cite{Fokas2016}, and has become a standard in the field for ensuring consistency, interpretability, and biological relevance in network-based protein analysis.

The resulting graph captures the topological architecture of the protein, allowing for further analysis such as centrality measures, clustering, and community detection. RINs are particularly useful for identifying functionally or structurally critical residues, often termed ``hotspots.” These may be allosteric sites, binding regions, or residues central to protein folding. Their detection can inform mutagenesis experiments, drug target identification, and the study of protein evolution \cite{doi:10.1073/pnas.0810961106}. In our approach, RINs serve as input to our quantum optimization framework, wherein the goal is to identify the top-$\tau$ central residues using eigenvector and Estrada QUBO formulations implemented through established tools in quantum annealing.

\subsection{Centrality Measures}

In the context of network analysis, centrality measures are used to identify nodes that hold strategic importance in terms of connectivity, communication, or influence \cite{Koschutzki2005-fn}. Several centrality measures have been applied to biological networks, each capturing different aspects of node importance. \textit{Degree centrality} counts how many direct connections a node has, identifying locally connected hubs within a protein’s RIN. \textit{Betweenness centrality} quantifies how often a node lies on shortest paths between pairs of nodes, using the fraction of those paths that pass through it \cite{353871a7-aafc-320a-b8a4-7a1ca0041e9d}. \textit{Closeness centrality} is the inverse of the average shortest distance from a node to all others, indicating its global reach in the network. The focus of our work is on \textit{eigenvector centrality}, a more inclusive metric that considers not just the number of connections but also the importance of the connected nodes. This implies that a node connected to highly central nodes is itself considered important. For example, in a protein RIN, a residue may not be highly connected itself but could interact with a tightly-knit cluster of functionally critical residues, thereby attaining a high eigenvector centrality score. Unlike degree or closeness centrality, eigenvector centrality captures the global topology of the network and aligns well with biological interpretations of cooperative behavior among residues \cite{Negre2018}.

\subsection{QUBO Formulation for Top-$\tau$ Eigenvector Centrality}

We begin by framing eigenvector centrality in network terms, where the centrality vector $x \in \mathbb{R}^n$ satisfies:
\[
A x = \lambda x
\]
with $\lambda$ as the principal eigenvalue of the adjacency matrix $A$ \cite{Bonacich1972-lm}. In residue interaction networks (RINs), the $i$-th component of $x$ represents the centrality score of residue $i$, capturing both direct interactions and the importance of connected neighbors. Eigenvector centrality highlights not only local hubs but also residues involved in globally influential clusters. This perspective aligns with how information propagates in protein structures. We recast this as an unconstrained quadratic optimization problem \cite{Lucas2014-ia}: selecting a binary vector $x \in \{0,1\}^n$ that identifies the top-$\tau$ central residues. Specifically, we aim to maximize the quadratic form $x^T Q x$ while enforcing that exactly $\tau$ nodes are selected. This leads to a QUBO (Quadratic Unconstrained Binary Optimization) formulation:
\[
Q = -P_0 A^2 \hat{d} \hat{d}^T A - P_0 A \hat{d} \hat{d}^T A^2 + P_1 C
\]
where $\hat{d} = d / \|d\|$ is the normalized degree vector, and $C$ enforces the $\tau$-node selection constraint
\[
C = (1 - 2\tau) I + U, \quad U_{ij} = 1 \text{ for } i \neq j
\]
where $P_0$ and $P_1$ have distinct roles: $P_0$ emphasizes selecting residues that are well-connected within the network (maximizing centrality), while $P_1$ ensures that exactly $\tau$ nodes are chosen by penalizing deviations from this target \cite{Akrobotu2022}.

In practice, $P_0$ is set to $1/\sqrt{n}$ and $P_1$ to a multiple of $n$, as these values effectively balance centrality maximization with constraint satisfaction. In our implementation, we employ a simpler form of the QUBO matrix
\[
Q = -P_0 A \hat{d} \hat{d}^T A - P_0 A \hat{d} \hat{d}^T A + P_1 C
\].
We use this form because it is numerically better behaved and easier for the QUBO solver to handle. By avoiding higher-order matrix powers (such as $A^2$), we reduce the risk of introducing large or small eigenvalues that can destabilize the optimization. This simpler form also minimizes numerical errors and ensures that the cardinality constraint remains effectively enforced. Consequently, it reliably identifies top-ranked residues without overshadowing the selection constraint or introducing unnecessary complexity into the energy landscape.

The parameters \( P_0 = \frac{1}{\sqrt{n}} \) and \( P_1 = 10n \) used in the QUBO formulation were chosen to balance the two competing objectives in the optimization landscape: maximizing spectral centrality (via \( P_0 \)) and enforcing the cardinality constraint of selecting exactly \( \tau \) residues (via \( P_1 \)). The scaling of \( P_0 \) with \( \frac{1}{\sqrt{n}} \) is motivated by results from continuous-time quantum walk theory, where this value represents the optimal transition amplitude for locating a marked node in symmetric quantum search problems~\cite{PhysRevA.70.022314, Akrobotu2022}. 

In our implementation, we found that this combination—\( P_0 = \frac{1}{\sqrt{n}} \) and \( P_1 = 10n \)—yielded the most stable and interpretable results across all tested protein networks. While Akrobotu \textit{et al.} reported that a smaller value of \( P_1 = 5n \) was sufficient for their synthetic and benchmark graphs, our findings suggest that the optimal ratio between \( P_1 \) and \( P_0 \) is sensitive to the structure and spectral properties of the graph, as well as the specific details of the QUBO formulation. This highlights a promising direction for future investigation: systematically exploring how penalty parameter scaling affects solution quality, interpretability, and robustness in biologically realistic networks. Such studies could inform the development of adaptive or data-driven strategies for parameter selection tailored to the structural diversity of protein graphs.

\subsection{Classical Computation of Eigenvector Centrality using NetworkX}

To establish a classical benchmark for evaluating quantum optimization results, we computed the eigenvector centrality of residues using the \texttt{NetworkX} library \cite{NetworkX}. NetworkX provides a built-in function \texttt{eigenvector\_centrality()}, which applies the power iteration method to approximate the leading eigenvector. In our implementation, we set the maximum number of iterations to 1000 and the convergence tolerance to $10^{-6}$ to ensure accurate and consistent results across different residue interaction networks. The function returns a dictionary mapping each node (residue) to its centrality score. These values are then sorted to identify the top-$\tau$ most central residues. For visualization, we scaled the node sizes in the protein-residue interaction graph according to their eigenvector centrality values. The resulting graph was rendered using \texttt{Matplotlib} and \texttt{NetworkX} \cite{NetworkX}, with larger nodes indicating higher centrality. This visual aid provides intuitive confirmation of the residues' relative importance within the protein structure. The use of NetworkX offers a fast and reliable classical reference for spectral centrality, enabling direct comparison with results from our QUBO-based quantum and simulated annealing formulations. It also serves as a fallback tool for validating quantum method output or analyzing convergence failures.

\subsection{Scaling and complexity}

The QUBO formulation provides computational advantages over conventional centrality methods. 
Eigenvector centrality, solved by power iteration, requires $O(mk)$ operations for a graph with $m$ edges, 
where the iteration count $k$ depends on the spectral gap $(\lambda_1 - \lambda_2)$ and the desired precision $\varepsilon$.
Estrada centrality involves computing a matrix exponential, which typically costs $O(n^3)$ for a network with $n$ residues. \cite{ESTRADA2000713}
In contrast, the QUBO objective
\[
f(x) = x^\top Q x,
\]
can be evaluated in time proportional to the number of nonzero entries in $Q$, i.e., $O(\mathrm{nnz}(Q)) \sim O(m)$, 
where $m$ is the number of edges in the residue interaction network.\cite{Akrobotu2022}
This reduction from cubic or iterative eigenvalue computations to linear scaling in the number of edges highlights the speedup potential. 
Moreover, heuristic solvers such as simulated annealing operate in sweeps of $O(m)$, while quantum annealing can further accelerate convergence 
by exploiting tunneling to escape local minima. 
Thus, the QUBO approach offers a scalable and hardware-compatible alternative for residue centrality prediction.

\section{Implementation}

\subsubsection{Data Preparation}

The first step in the workflow is to prepare structural data for constructing the residue interaction network (RIN). Protein structures are obtained from the Protein Data Bank (PDB)\cite{10.1093/nar/28.1.235}, which provides three-dimensional atomic coordinates. For instance, in our case study, we use the oxytocin protein with PDB ID \texttt{1XY1}. We fetch the corresponding PDB file using Biopython’s \texttt{PDBList} module and parse the structure with \texttt{Bio.PDB}\cite{Biopython}. From the full structure, we extract the coordinates of the C$\alpha$ atoms, which serve as coarse-grained representations of residues. Residue pairs are then connected by edges if their C$\alpha$ atoms lie within a distance cutoff of 8.0~\AA, a threshold widely used to capture relevant interactions.\cite{Fokas2016} This process yields a graph-based model where nodes correspond to residues and edges represent significant spatial contacts.

\subsubsection{Graph Construction and Visualization}

Using the \texttt{NetworkX} library, we construct an undirected, unweighted graph $G=(V,E)$ from the identified residue pairs. Each residue is modeled as a node indexed by its sequence position, and edges encode residue-residue contacts within the defined distance threshold. The resulting RIN preserves the protein’s structural topology, capturing both local and global connectivity. For visualization, we use \texttt{NetworkX} and \texttt{Matplotlib} to create network plots that show how residues are connected. The layout arranges nodes to reflect their distances in the network. We label and color nodes by their centrality scores, helping us spot important residues and understand the overall structure of the protein.

\subsubsection{Adjacency Matrix and Centrality Computation}

The RIN is encoded numerically as an adjacency matrix $A$, where $A_{ij}=1$ if residues $i$ and $j$ are connected, and 0 otherwise. We generate this binary adjacency matrix using NetworkX and store it for further analysis, including both classical and quantum algorithms \cite{Hagberg2008-wq}.

\subsubsection{Simulated Annealing}

To benchmark the performance of our quantum-inspired approach, we implemented a classical simulated annealing (SA) method \cite{Kirkpatrick1983-yw} using D-Wave’s \texttt{SimulatedAnnealingSampler}\cite{DwaveOcean} from the \texttt{dimod} framework. SA is a stochastic optimization algorithm that emulates the process of slowly cooling a system to reach its ground state by minimizing the energy function:
\[
E(x) = x^T Q x
\]
where $Q$ is the QUBO matrix derived from the adjacency matrix $A$ and normalized degree vector $\hat{d}$. The temperature parameter $\beta$ is gradually increased to encourage convergence to low-energy solutions. The QUBO matrix is converted into a Binary Quadratic Model (BQM) using \texttt{dimod.from\_numpy\_matrix()}, and the SA sampler explores binary solutions over 10{,}000 reads, with $\beta$ ranging from 0.1 to 4.0. Among these samples, we retain only those solutions that satisfy the constraint $\sum_i x_i = \tau$, selecting the configuration with the lowest energy as the optimal residue set. D-Wave’s simulated annealing mimics physical thermalization processes and serves as a classical precursor to its quantum annealing hardware \cite{RevModPhys.80.1061}. In practice, this classical SA approach consistently identified top-$\tau$ residues that closely matched classical eigenvector centrality rankings.

\begin{figure}[htbp]
    \centering
    \begin{tikzpicture}[
        node distance=1cm,
        box/.style={
            rectangle, draw=black, rounded corners,
            text centered, text width=3.7cm, minimum height=1cm, fill=blue!10
        },
        arrow/.style={->, thick}
    ]

    \node[box] (pdb) {Download structure\\ from PDB};
    \node[box, below=of pdb] (rin) {Construct residue\\ interaction network};
    \node[box, below=of rin] (qubo) {Formulate QUBO matrix};
    \node[box, below=of qubo] (sa) {Annealing};
    \node[box, below=of sa] (topresidues) {Predict top-$\tau$ central residues};

    \draw[arrow] (pdb) -- (rin);
    \draw[arrow] (rin) -- (qubo);
    \draw[arrow] (qubo) -- (sa);
    \draw[arrow] (sa) -- (topresidues);

    \end{tikzpicture}
    \caption{
        \textbf{Schematic workflow of the RinQ pipeline.}}

    \label{fig:RinQ_pipeline}
\end{figure}
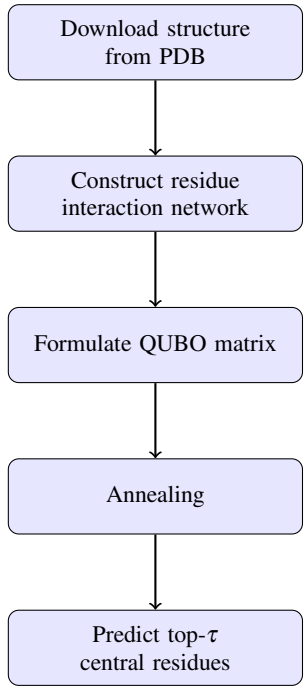

\subsection{Estrada Centrality}

Estrada centrality measures how easily information can flow to and from a residue by considering all possible walks that start and end at that residue. It is calculated as the diagonal entries of the matrix exponential of the adjacency matrix:
\[
\text{Estrada}(i) = \left[\exp(A)\right]_{ii}
\]
where $A$ is the adjacency matrix and $\exp(A)$ sums up all possible walks in the network, giving shorter walks more weight because of the factorial term in the series expansion \cite{PhysRevE.82.066102}.

\subsubsection{QUBO Formulation for Top-$\tau$ Estrada Nodes}

To extend our quantum optimization framework beyond eigenvector centrality, we formulate the problem of identifying the top-$\tau$ Estrada central nodes as a QUBO problem. We start by defining a truncated approximation of the matrix exponential $\exp(A)$:
\[
E = I + A + 0.5A^2 + \frac{1}{6}A^3
\]
where $A$ is the adjacency matrix. It should be noted that this truncated matrix exponential approach is an approximation: while it reduces computational complexity and numerical challenges associated with evaluating the full matrix exponential, it may also limit the precision of the Estrada centrality calculations, particularly for networks with complex topologies.

To quantify the approximation error introduced by truncating the matrix exponential $\exp(A)$ at the cubic term $A^3$, we consider the Frobenius norm of the remainder series. The matrix exponential admits the Taylor expansion $\exp(A) = \sum_{k=0}^{\infty} \frac{A^k}{k!}$, and truncating this at $k=3$ yields a remainder $\epsilon = \sum_{k=4}^{\infty} \frac{A^k}{k!}$. Taking the Frobenius norm gives a global upper bound on the error:
\[
\|\epsilon\|_F \leq \sum_{k=4}^{\infty} \frac{\|A\|_F^k}{k!}.
\]
For the adjacency matrix $A$ of an undirected residue interaction network (RIN), where $A_{ij} = 1$ if residues $i$ and $j$ interact and $0$ otherwise, and with no self-loops ($A_{ii} = 0$), the Frobenius norm satisfies
\[
\|A\|_F = \sqrt{\sum_{i=1}^n \sum_{j=1}^n A_{ij}^2} = \sqrt{2|E|},
\]
since each undirected edge $\{i,j\}$ contributes two nonzero entries to $A$ (i.e., $A_{ij}$ and $A_{ji}$). Substituting into the bound, we obtain
\[
\|\epsilon\|_F \leq \sum_{k=4}^{\infty} \frac{( \sqrt{2|E|} )^k}{k!}.
\]
This result illustrates that the truncation error scales nonlinearly with the number of edges $|E|$. While the error is negligible for sparse graphs, it grows rapidly for denser networks—potentially compromising the accuracy of Estrada centrality computations in densely connected proteins.

We then compute the degree-normalized vector $\hat{d}$ and form the following low-rank matrix:
\[
E \cdot (\hat{d} \hat{d}^T) \cdot E
\]
This matrix captures the influence of nodes through the low-rank structure of the matrix exponential. The final QUBO matrix is
\[
Q = -P_0 \cdot (E \cdot (\hat{d} \hat{d}^T) \cdot E) + P_1 \cdot C,
\]
where $P_0 = \frac{1}{\sqrt{n}}$, $P_1 = 10n$, and $C = (1 - 2\tau)I + U$ enforces the selection of exactly $\tau$ residues. 

\subsubsection{NetworkX Implementation of Estrada Centrality}

We also created a classical baseline for Estrada centrality using \texttt{NetworkX} and \texttt{scipy}'s \texttt{expm} function. The adjacency matrix $A$ is extracted from the protein-residue interaction network $G$, and the matrix exponential $\exp(A)$ is calculated directly. We use the diagonal entries of $\exp(A)$ as the Estrada centrality scores. These scores are saved in a dictionary linking each residue (node) to its score. We rank the residues in descending order to find the top-$\tau$ most central nodes. Finally, we visualize the protein graph with node sizes scaled to Estrada scores, helping us see how communicability is distributed across residues.

\subsubsection{Simulated Annealing of QUBO-Estrada Formulation}

To solve the Estrada-based QUBO problem using classical quantum-inspired methods, we employed D-Wave's \texttt{SimulatedAnnealingSampler}. The QUBO matrix constructed from the projected Estrada kernel is first converted into a \texttt{BinaryQuadraticModel} object. We then perform annealing with 10,000 reads and a temperature range from $\beta = 0.1$ to $\beta = 4.0$. Post-processing involves filtering solutions to retain only those binary vectors that select exactly $\tau$ nodes. Among these, the solution with the lowest QUBO energy is chosen as the optimal configuration. The resulting bitstring identifies the top-$\tau$ residues with the highest Estrada influence under the constraints of the QUBO model. The simulated annealing approach consistently finds valid solutions across all tested proteins, serving as a strong baseline. Importantly, the solutions from simulated annealing closely resemble those from quantum annealers, making it a useful way to mimic the behavior of real quantum hardware \cite{McGeoch2013-yx}.

\section{Results and Analysis}

\subsection{Proteins Tested}

We selected a representative set of proteins and peptides, ranging from small structural motifs to functionally diverse biomolecules\cite{Dyson2005-aa}, to benchmark the performance of our approach. The following are the specific structures examined in this study:

\begin{itemize}
    \item \textbf{1A7F – Insulin mutant (29 residues):} A peptide hormone that regulates the endocrine system and inhibits the secretion of several other hormones.
    
    \item \textbf{1GCN – Glucagon (29 residues):} A metabolic hormone that raises blood glucose levels by promoting gluconeogenesis and glycogenolysis.

    \item \textbf{1JL9 – EGF (Epidermal Growth Factor) (51 residues):} A human growth factor involved in the stimulation of cell growth, proliferation, and differentiation.

    \item \textbf{1Q71 – microcin J25 (21 residues):} A 21-amino acid lasso peptide that inhibits the growth of Gram-negative bacteria by targeting RNA polymerase.

    \item \textbf{1UBQ – Ubiquitin (76 residues):} A highly conserved regulatory protein that tags other proteins for degradation via the proteasome pathway.

    \item \textbf{1XY1 – Deamino-Oxytocin (10 residues):} A synthetic analog of oxytocin, a neuropeptide associated with labor, lactation, and social bonding.

    \item \textbf{2K6O – LL-37 (37 residues):} An antimicrobial peptide from the innate immune system with broad-spectrum activity against pathogens.

    \item \textbf{2MLT – Melittin (Bee Toxin) (27 residues):} The major active component in bee venom, known for its membrane-lytic and antimicrobial properties.

    \item \textbf{2N08 – Small Peptide (12 residues):} A minimal peptide model used for analyzing structural motifs and basic residue interactions.

    \item \textbf{4D5M – Small Peptide Fragment (11 residues):} A fragment of a larger protein, providing simplified structural context for residue-level analysis.

    \item \textbf{6A5J – Small Peptide Fragment (13 residues):} A peptide fragment studied for its relevance in peptide–receptor binding and flexibility.

    \item \textbf{6RQS – Small Peptide (18 residues):} A small peptide with known structure, serving as a minimal model for folding and stability studies.
\end{itemize}

The network visualizations of all the proteins are available in supplementary material.
\subsection{Case Study: Oxytocin (1XY1)}

In this case study, we focus on the Oxytocin protein \cite{Gimpl2001-sd}(PDB ID: 1XY1) to illustrate how our framework integrates classical and quantum approaches to identify the most central residues within a residue interaction network. While a complete list of all tested proteins is presented in Table~\ref{tab:centrality-comparison}, we selected 1XY1 as a representative example to provide detailed visualizations of the RIN scaled by different centrality measures and to demonstrate how the various computational elements of our pipeline converge in practice.

\subsubsection{Eigenvector Centrality Analysis}

The top residues ranked by classical eigenvector centrality are as follows:

\begin{itemize}
    \item Residue 6: 0.4545
    \item Residue 5: 0.3876
    \item Residue 1: 0.3820
    \item Residue 2: 0.3402
    \item Residue 3: 0.3402
    \item Residue 7: 0.3049
    \item Residue 4: 0.2807
    \item Residue 8: 0.2446
    \item Residue 9: 0.1851
\end{itemize}

Using the QUBO-based quantum optimization approach, the best valid sample identified the following top nodes:

\begin{itemize}
    \item Best valid sample: \{0: 1, 1: 1, 2: 1, 3: 0, 4: 1, 5: 1, 6: 0, 7: 0, 8: 0\}
    \item Energy: -2474.5714285714284
    \item Top nodes: [1, 2, 3, 5, 6]
\end{itemize}

\subsubsection{Estrada Centrality Analysis}

The top residues ranked by classical Estrada centrality are as follows:

\begin{itemize}
    \item Residue 6: 47.1546
    \item Residue 5: 34.7512
    \item Residue 1: 33.6590
    \item Residue 3: 27.4557
    \item Residue 2: 27.4557
    \item Residue 7: 22.8833
    \item Residue 4: 19.3854
    \item Residue 8: 15.8201
    \item Residue 9: 10.0360
\end{itemize}

The QUBO-based approach for Estrada centrality identified the following:

\begin{itemize}
    \item Best valid sample: \{0: 0, 1: 0, 2: 0, 3: 0, 4: 0, 5: 1, 6: 0, 7: 0, 8: 0\}
    \item Energy: -606.5291005290992
    \item Top nodes (Estrada centrality): [6]
\end{itemize}

These results highlight the ability of the quantum optimization framework to pinpoint central residues in protein networks, complementing classical methods and providing new insights into residue importance.

\subsubsection{Visualization of central residues}

\begin{figure}[H]
    \centering
    \includegraphics[width=0.90\columnwidth]{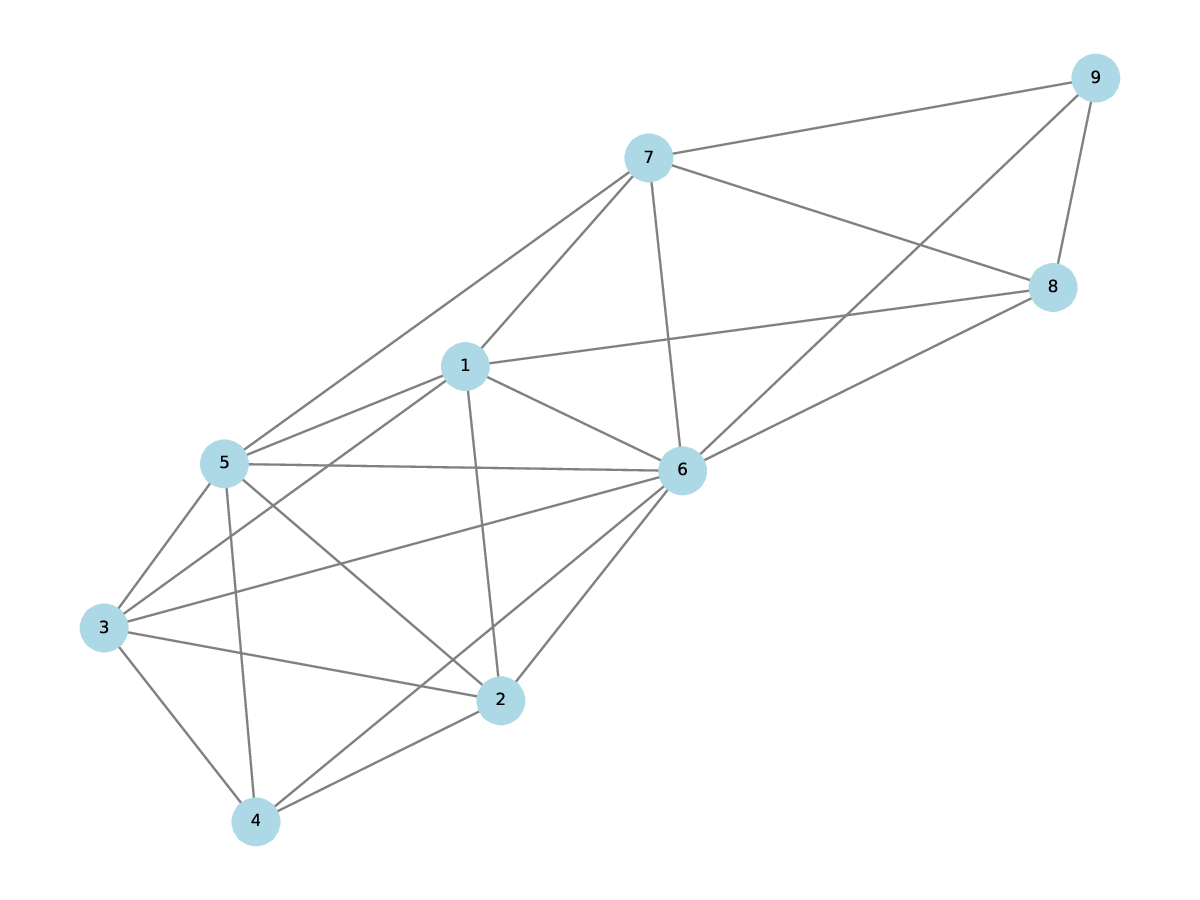}
    \caption{Residue Interaction Network of 1XY1. Nodes represent residues and edges denote interactions within the specified cutoff distance.}
    \label{fig:basic_network}
\end{figure}
\vspace{-2em} 

\begin{figure}[H]
    \centering
    \includegraphics[width=0.90\columnwidth]{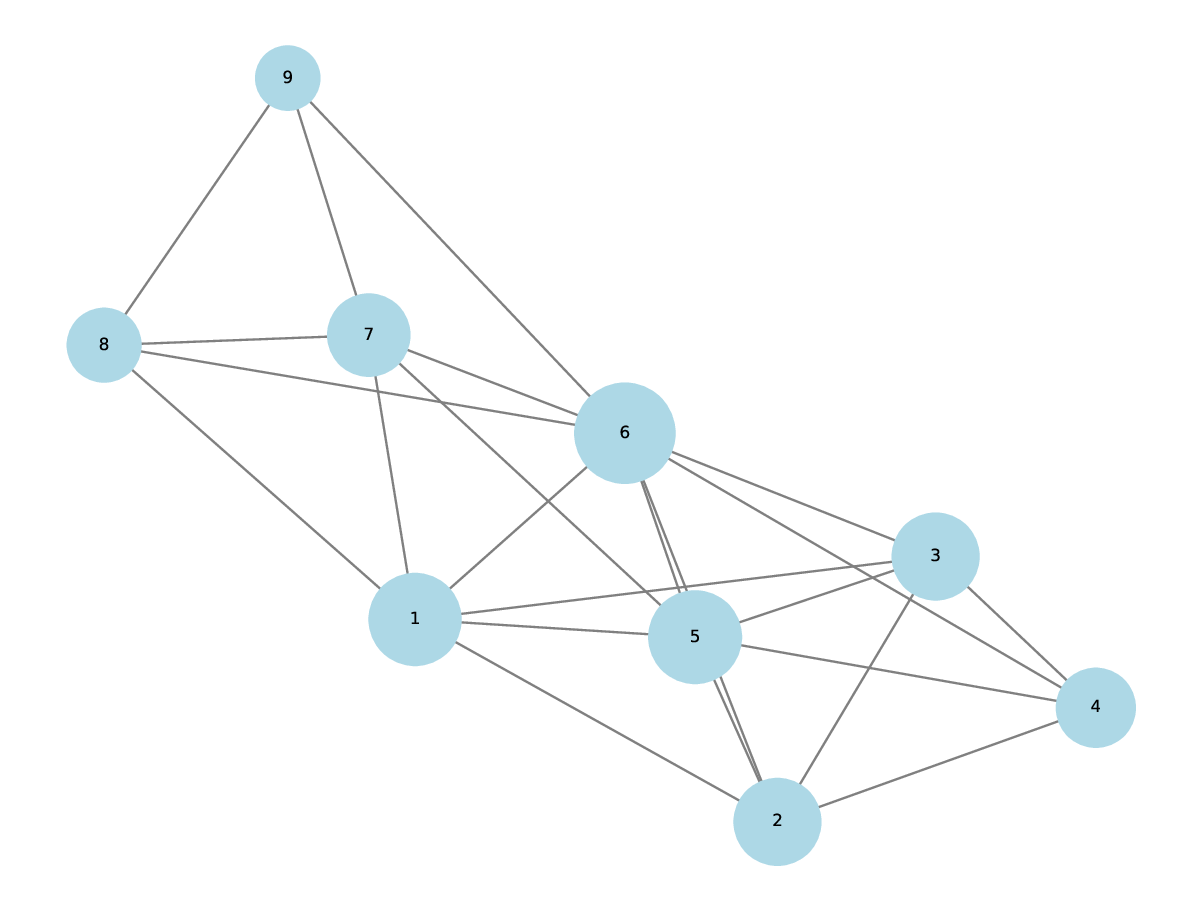}
    \caption{Residue Interaction Network of 1XY1 highlighting eigenvector centrality. Node sizes are scaled according to their eigenvector centrality values.}
    \label{fig:eigenvector_centrality}
\end{figure}
\vspace{-3em} 

\begin{figure}[H]
    \centering
    \includegraphics[width=0.90\columnwidth]{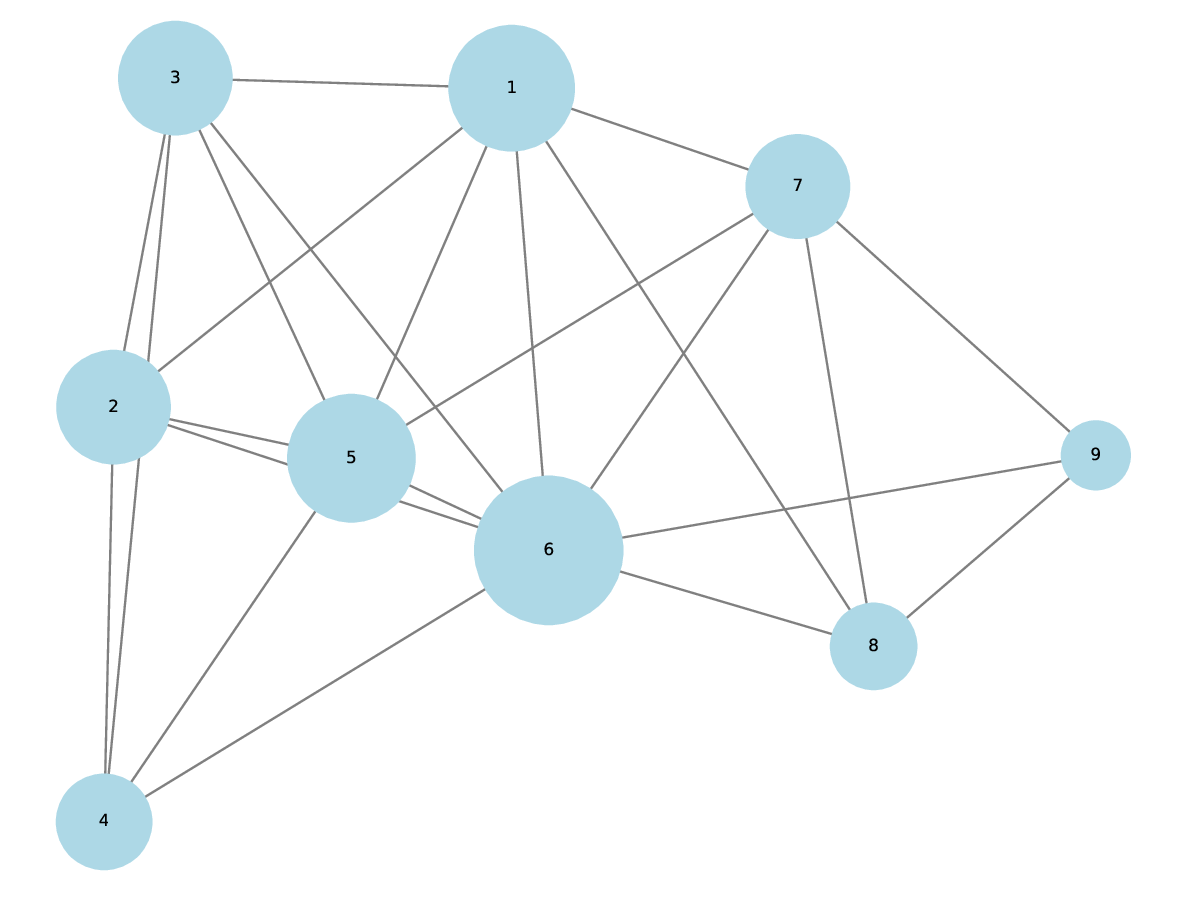}
    \caption{Residue Interaction Network of 1XY1 highlighting Estrada centrality. Node sizes are scaled by normalized Estrada centrality values.}
    \label{fig:estrada_centrality}
\end{figure}

\subsubsection{Experimental Validation for Oxytocin (1XY1).}
In this study, the QUBO-based eigenvector centrality method identified Tyr$^2$, Ile$^3$, Asn$^5$, Cys$^6$, as the most central residues in oxytocin (PDB ID: 1XY1). Notably, several of these residues—Tyr$^2$, Ile$^3$, and Cys$^6$—have been independently validated in the literature as critical for oxytocin's structural stability and receptor binding. Tyr$^2$ and Cys$^6$ were identified as key ligand residues in receptor activation~\cite{Liu}, while molecular dynamics and ion-mobility mass spectrometry studies showed that Tyr$^2$, Ile$^3$, and Cys$^6$ participate in Zn$^{2+}$ coordination and exhibit restricted mobility, indicative of structural or functional importance~\cite{md_ot_zinc, ims_ot_bd}. The overlap between RinQ-predicted hotspots and experimentally established functional sites highlights the biological validity of the QUBO-based approach and supports its use in functional residue identification.

\begin{table*}[ht]
\centering
\caption{Top residues (nodes) for each protein based on Eigenvector centrality, QUBO-based Eigenvector centrality, and their Jaccard Index agreement.}
\label{tab:centrality-comparison}
\begin{tabular}{|l|p{3.8cm}|p{3.8cm}|c|}
\hline
\textbf{Protein (PDB ID)} & \textbf{Eigenvector Centrality} & \textbf{QUBO Eigenvector Centrality} & \textbf{Jaccard Index} \\
\hline
1A7F  & 11, 6, 12, 14, 10         & 11, 14, 15, 16, 17         & 0.250 \\
1GCN  & 17, 18, 16, 19, 15        & 16, 17, 18, 19, 20         & 0.667 \\
1JL9  & 14, 42, 33, 31, 18        & 26, 28, 36, 37, 40         & 0.000 \\
1Q71  & 19, 18, 7, 8, 20          & 7, 8, 16, 18, 19           & 0.667 \\
1UBQ  & 5, 67, 4, 3, 69           & 66, 67, 68, 69, 70         & 0.250 \\
1XY1  & 6, 5, 1, 2, 3             & 1, 2, 3, 5, 6              & 1.000 \\
2K6O  & 18, 13, 16, 17, 15        & 18, 19, 21, 26, 29         & 0.111 \\
2MLT  & 12, 13, 9, 14, 11         & 8, 9, 11, 12, 14           & 0.667 \\
2N08  & 6, 5, 7, 8, 4             & 4, 5, 6, 7, 8              & 1.000 \\
4D5M  & 5, 3, 6, 4, 2             & 3, 4, 5, 6, 8              & 0.667 \\
6A5J  & 8, 9, 6, 7, 5             & 5, 6, 7, 8, 9              & 1.000 \\
6RQS  & 10, 11, 9, 8, 7           & 7, 8, 9, 10, 11            & 1.000 \\
\hline
\end{tabular}
\end{table*}

\begin{table*}[ht]
\centering
\caption{Top residues (nodes) for each protein based on Estrada centrality and QUBO-based Estrada centrality.}
\label{tab:estrada-comparison}
\begin{tabular}{|l|p{4cm}|p{4cm}|}
\hline
\textbf{Protein (PDB ID)} & \textbf{Estrada Centrality} & \textbf{QUBO Estrada Centrality} \\
\hline
1A7F & 11, 6, 12, 14, 10 & 25 \\
1GCN & 23, 19, 20, 18, 17 & 17 \\
1JL9 & 14, 42, 33, 31, 18 & 37 \\
1Q71 & 19, 18, 7, 8, 20 & 15 \\
1UBQ & 5, 4, 23, 3, 6 & 6 \\
1XY1 & 6, 5, 1, 3, 2 & 6 \\
2K6O & 29, 13, 10, 18, 14 & 29 \\
2MLT & 12, 9, 13, 8, 14 & 14 \\
2N08 & 6, 5, 7, 8, 4 & 8 \\
4D5M & 5, 3, 6, 8, 4 & 5 \\
6A5J & 8, 9, 6, 7, 5 & 8 \\
6RQS & 10, 11, 9, 8, 12 & 10 \\
\hline
\end{tabular}
\end{table*}

\begin{table}[h]
\centering
\caption{Comparison of NetworkX EC Scores and QUBO Rankings for 1XY1, 6RQS, 6A5J, and 2N08}
\label{tab:ranking-centrality}
\begin{tabular}{|c|c|c|c|}
\hline
\textbf{Protein} & \textbf{Residue} & \textbf{NetworkX EC Score} & \textbf{QUBO EC Rank} \\
\hline
\multirow{5}{*}{1XY1} 
& 6 & 0.4545 & 1 \\
& 5 & 0.3876 & 2 \\
& 1 & 0.3820 & 3 \\
& 2 & 0.3402 & 4 \\
& 3 & 0.3402 & 5 \\
\hline
\multirow{5}{*}{6RQS}
& 10 & 0.3359 & 1 \\
& 11 & 0.3246 & 2 \\
& 9  & 0.3135 & 3 \\
& 8  & 0.3078 & 4 \\
& 7  & 0.2934 & 5 \\
\hline
\multirow{5}{*}{6A5J}
& 8 & 0.3586 & 1 \\
& 9 & 0.3553 & 2 \\
& 6 & 0.3395 & 3 \\
& 7 & 0.3297 & 4 \\
& 5 & 0.3212 & 5 \\
\hline
\multirow{5}{*}{2N08}
& 6 & 0.4046 & 1 \\
& 5 & 0.3760 & 2 \\
& 7 & 0.3488 & 3 \\
& 8 & 0.3155 & 4 \\
& 4 & 0.3070 & 5 \\
\hline
\end{tabular}
\end{table}

\subsection{Analysis of Results}

Table~\ref{tab:centrality-comparison} and Table~\ref{tab:estrada-comparison} summarizes the top residues identified by each centrality measure across the tested proteins, including classical eigenvector and Estrada centralities, as well as their QUBO-based counterparts. For centrality methods, the top 5 residues are reported, except for the QUBO-based Estrada centrality, which, due to the current formulation and penalty constraints, typically yields only the top residue (or up to the top 2--3 residues). Here, as a proof-of-exercise, we present only the top-ranked residue, leaving further refinement of the QUBO formulation and penalty functions as an area for future work. Overall, there is strong agreement between classical and QUBO-based methods, particularly for eigenvector centrality, with minor discrepancies arising in cases like 1Q71 and 2K6O. These discrepancies are likely due to the stochastic nature of simulated annealing and the sensitivity of QUBO solutions to penalty constraints.

We observe that for small, symmetric, and compact peptides such as \textbf{1XY1}, \textbf{2N08}, \textbf{6A5J}, and \textbf{6RQS}---all of which contain fewer than 20 residues and exhibit relatively regular residue--residue interaction graphs---the QUBO-based approach achieves perfect agreement with classical results (\textit{Jaccard Index} = 1.000).

However, for larger or structurally asymmetric proteins, such as \textbf{1UBQ} (76 residues), \textbf{1JL9} (51 residues), and \textbf{2K6O} (37 residues), the Jaccard index is significantly lower, with some cases (e.g., \textbf{1JL9}) exhibiting no overlap in the top-5 central residues. This discrepancy is expected: the original QUBO-based formulation was designed and validated on synthetic or regular graphs \cite{Akrobotu2022}, such as complete graphs and expanders, where node degrees are homogeneous and eigenvector alignment is more tractable. In contrast, protein residue interaction networks (RINs) are highly irregular, sparse, and topologically diverse, often exhibiting domain-specific connectivity, long-range couplings, and local clustering.

This intrinsic asymmetry contributes to slight irregularities in the QUBO-based results for larger proteins and peptides, though smaller peptides tend to show excellent agreement with classical NetworkX results. These insights highlight both the promise and challenges of applying quantum-inspired methods to real-world biological networks. For larger proteins, stronger penalty constraints are often required to enforce the $\tau$-residue selection condition, highlighting the need for continued refinement of the QUBO formulation. Nevertheless, this prototype demonstrates that our approach can reliably identify central residues. It lays a strong foundation for applying quantum methods to centrality detection in residue interaction networks.

As shown in Table~\ref{tab:ranking-centrality}, we have extended our analysis by exploring a range of $\tau$ values from 1 to 5 for those proteins where the QUBO-based eigenvector centrality perfectly matched the classical ranking (i.e., Jaccard Index = 1.000). These proteins—\textbf{1XY1}, \textbf{2N08}, \textbf{6A5J}, and \textbf{6RQS}—exhibited highly regular and compact residue interaction graphs, where we are most confident in the QUBO formulation’s predictive accuracy. The QUBO-based rankings of residues were determined by sweeping $\tau$ from 1 to 5, allowing us to incrementally construct the top-5 central residues for each protein. This approach provides a consistent ranking profile and enables direct comparison with classical eigenvector centrality results.

\section{Biological Relevance of Predicted Central Residues}

Identifying central residues in protein structures is not only of theoretical interest but also has important biological implications \cite{doi:10.1073/pnas.0810961106}. Numerous studies have demonstrated that residues occupying central positions in residue interaction networks (RINs) frequently coincide with functional sites, such as catalytic residues, allosteric communication hubs, and binding interfaces \cite{Yang2009-mh, Konno2019-xe, 10.1093/nar/gkn433}. Numerous studies have shown that centrality measures align well with known functional sites in proteins. For example, eigenvector centrality has been found to correlate with experimentally validated allosteric pathways, identifying residues critical for communication and regulation within protein structures \cite{Reetz2009-dm, Negre2018}. 

Estrada centrality captures how residues contribute to the overall folding and compactness of protein structures, key for biological function.\cite{ESTRADA2000713} By weighing short walks more heavily, it highlights residues that stabilize the protein’s 3D shape and enable efficient communication pathways. This makes it especially relevant for pinpointing residues involved in allosteric signaling, and structural integrity. More generally, centrality metrics such as closeness and betweenness have been shown to identify catalytic residues and residues involved in conformational changes \cite{Chea2007-ud, Negre2018, ESTRADA2000713}. These findings reinforce the biological relevance of centrality-based rankings and underscore the potential of computational approaches to guide experimental validation and functional interpretation of protein structures. \cite{Kovacs2005-km}

\section{Discussion}

The QUBO-based formulation developed in this work offers a powerful and generalizable method for identifying the most central residues in residue interaction networks. Its key strength lies in the ability to encode a centrality objective—such as eigenvector or Estrada centrality—within a binary optimization framework suitable for both classical and quantum solvers. A key advantage of this design is the flexibility offered by the parameters $\tau$, $P_0$, and $P_1$. The parameter $\tau$ sets the number of top residues to select, while $P_0$ determines how strongly we emphasize centrality and $P_1$ ensures that exactly $\tau$ residues are chosen. In our tests, values such as $P_0 = \frac{1}{\sqrt{n}}$ and $P_1 = 10n$ yielded consistent and robust results for the smaller proteins.

The formulation is not without limitations, with one key challenge is scalability: as protein size increases, the adjacency matrix—and consequently the QUBO matrix—grows quadratically in dimension \cite{Hauke_2020}. This poses practical barriers for existing quantum hardware, which are currently limited to a few dozen qubits at best. Furthermore, the QUBO formulation inherently returns a bitstring corresponding to a binary decision vector (select vs. not select), but it does not impose a strict ranking among the top-$\tau$ nodes. This can lead to degeneracy in solutions, especially in graphs with high symmetry or redundant topology, where multiple subsets yield similar optimization scores. The current framework also depends on access to quantum hardware, which is limited in availability and subject to queue times and noise. 

While simulations and classical solvers like D-Wave's simulated annealing are viable stand-ins, they do not replicate the quantum sampling behavior expected from gate-based or annealing-based quantum devices. Moreover, QUBO performance may degrade when applied to irregular or highly fragmented networks, where spectral measures become less meaningful due to disconnected components or weakly structured neighborhoods \cite{doi:10.1137/090761070}. Despite these challenges, the results remain promising and suggest that QUBO-based optimization has a valuable role to play in quantum-enhanced bioinformatics.  Techniques from quantum information processing are rapidly emerging as powerful tools for investigating soft matter systems such as proteins, with recent efforts advancing algorithmic strategies, encoding schemes, and representational frameworks that address the structural complexity, dynamical richness, and many-body interactions inherent to biologically relevant systems \cite{kumaran25, haghshenas25, glass25, activeglassy24,  khatami22}. Continued improvements in hardware accessibility, circuit design, and post-processing heuristics are expected to further improve solution quality and allow the method to scale to more complex and biologically realistic protein systems \cite{marthaler25, contQFIdr25, aplQ, oliveira2023fluctuation, drive22}.

\section{Future Work}

This work establishes a QUBO-based quantum optimization framework for identifying central residues in protein structures. While the results demonstrate promise, several research directions remain to be explored to fully realize the potential of quantum-enhanced protein network analysis. 

First, one of the most immediate goals is to execute the QUBO formulations on real quantum hardware, deployment of which will allow us to evaluate the influence of hardware noise, connectivity constraints, and decoherence on solution quality. Benchmarking against simulated annealing and other classical baselines will help quantify any quantum advantage and guide future algorithmic improvements.

Second, while we have already extended our framework to include Estrada centrality, future work could focus on incorporating other commonly used centrality measures, such as closeness, and betweenness. These alternative metrics may capture different aspects of residue importance in protein networks, particularly in cases where eigenvector or Estrada centralities are less informative. Exploring multiple centralities in parallel could lead to more robust predictions of functional or allosteric hotspots. A third major direction is to extend the model to dynamic or time-dependent residue interaction networks. Proteins are not static entities; they undergo conformational fluctuations and dynamic rearrangements \cite{Karplus2005-su}, particularly in the context of binding events or enzymatic activity. Incorporating structural ensembles from time-resolved crystallography or molecular dynamics simulations will enable a richer, temporally aware view of residue importance. A dynamic QUBO formulation may help identify residues that act as transient hubs or toggle between functional states.

Another obvious area of work involves all-$\tau$ analysis, where the QUBO is solved for a range of $\tau$ values to extract hierarchies of central residues rather than a fixed number. This would allow us to reconstruct full ranking profiles across the protein and to identify degenerate or near-degenerate configurations, offering deeper insights into structural redundancy or robustness. From a biological standpoint, a promising avenue for future work would involve either conducting experimental validation of the predicted central residues or comparing our predictions to existing experimental data, such as known active sites, ligand-binding regions, or mutationally critical residues. This validation effort could follow a more extensive benchmarking study across diverse proteins using our QUBO formulation and also leveraging real quantum devices. 

Another important direction is to enhance the predictive depth of the QUBO--Estrada formulation. In its current form, the method reliably identifies only the top one or two central residues, a limitation that may arise from penalty constraints in the QUBO encoding, approximation of the matrix exponential, or the stochastic nature of the optimization process. Future research could focus on refining the Hamiltonian representation to better capture communicability, incorporating error-mitigated quantum sampling strategies, and developing hybrid classical--quantum approaches that improve scalability while preserving accuracy. These refinements would help extend RinQ's applicability beyond the most dominant residues and enable a more nuanced identification of functionally relevant sites.

Longer term goals involve integrating centrality-based predictions into automated pipelines for drug discovery and protein engineering. By identifying structurally and functionally critical residues using quantum computing, our framework could aid in prioritizing mutagenesis experiments, guiding rational drug design, or informing synthetic biology applications. As quantum hardware continues to improve, we anticipate that quantum-enhanced residue analysis will become a key component in the structural bioinformatics toolbox.

\section{Conclusion}

RinQ establishes a scalable and biologically interpretable framework for integrating noisy intermediate-scale quantum (NISQ) devices into protein network analysis. While the present implementation demonstrates feasibility on small proteins, extending the approach to larger, structurally complex systems will require addressing key challenges, including scalability of the QUBO formulation, noise resilience on real quantum hardware, and adaptive parameter tuning. An additional limitation is that the present analysis is based on static protein structures from the Protein Data Bank (PDB). In physiological conditions, proteins undergo conformational dynamics driven by ligand binding, allosteric regulation, and post-translational modifications. Capturing these effects will require future extensions to dynamic or ensemble-based residue interaction networks, potentially informed by molecular dynamics simulations or time-resolved experimental data. 

Beyond these limitations, RinQ provides a foundation for incorporating additional centrality measures, dynamic structural ensembles, and experimental validation. The broader implication is that quantum optimization can serve as a practical tool for functional residue identification, with potential applications in mutagenesis studies, drug discovery, and protein engineering. More broadly, this work contributes to the growing intersection of quantum information science and structural biology, where quantum algorithms may transition from conceptual promise to tangible impact in molecular biosciences.

\section*{Data and Code Availability}

All data and code associated with this paper are available at the following GitHub repository: \url{https://github.com/IshmamShah/RinQ}

\section*{Acknowledgments}
The author acknowledges support from DoraHacks through the NISQ Quantum Grant.

\nocite{*}

%

\end{document}